\documentclass{appolb}
\usepackage{graphicx}
\begin{document}
\title{Photon-induced neutron, proton and alpha evaporation from heavy nucleus%
\thanks{Presented at the 57th Zakopane Conference on Nuclear Physics, {\it Extremes of the Nuclear Landscape}, Zakopane, Poland, 25 August–1 September, 2024.}%
}
\author{P. Jucha, K. Mazurek, M. Kłusek-Gawenda, M. Ciemała
\address{Institute of Nuclear Physics Polish Academy of Sciences,\\ PL-31342 Krak\'ow, Poland}
\\[2mm]
{A. Szczurek
\address{Institute of Nuclear Physics Polish Academy of Sciences,\\ PL-31342 Krak\'ow, Poland;\\ College of Mathematics and Natural Sciences, University of Rzeszów, \\PL-35959 Rzeszów, Poland}
}\\[2mm]
{Yuliia Shevchuk
\address{The Ivan Franko National University of Lviv,\\ 1, Universytetska str., Lviv, 79000, Ukraine}
}\\[2mm]
{S. Słotwi\'nski
\address{AGH University of Krakow\\
al. Adama Mickiewicza 30
PL-30059 Krak\'ow, Poland}
}
}

\maketitle
\begin{abstract}
In ultraperipheral heavy-ion collisions (UPCs) at the Large Hadron Collider (LHC) Pb nuclei are excited through interactions induced by strong electromagnetic fields. The expected excitation energy  could reach hundreds~MeV, which leads to the subsequent emission of various particles, including neutrons, protons, and alpha particles.

To accurately describe deexcitation of nuclei, we have developed two novel approaches. The first method utilizes the Heavy Ion Phase Space Exploration (HIPSE) model to simulate pre-equilibrium emissions and to estimate the excitation energy of the remaining nucleus. Our second approach introduces a new technique for modeling the excitation energy of the nucleus. This method consists of a two-component function to represent the excitation energy distribution more precisely, accounting for the energy loss due to the interaction between photons and quasi-deuteron. Both of these modeling techniques are integrated with the results produced by the GEMINI++ generator, which implements the Hauser-Feshbach formalism to simulate the statistical decay of excited nuclei. The alternative calculations are done with EMPIRE platform. Using these approaches, we obtained the cross sections for the emission of neutrons, protons, and alpha particles resulting from UPC at the LHC. 

Our results were compared with the experimental data of the ALICE group on neutron and proton multiplicities.

\end{abstract}
  
\section{Introduction}
The ultrarelativistic heavy ions are a source of strong electromagnetic fields. In the relativistic heavy ion collisions fast moving charges are the source of electromagnetic field, which can generate quasi-real photons. In ultraperipheral collisions (UPC) of ions, the nucleus interacts with the field, resulting in a Coulomb excitation of the ion \cite{Klusek-Gawenda:2013ema}. As an outcome, the photon-induced excitation leads to the emission of particles and nuclear fission. So far, the ALICE experiment has detected evaporated neutrons \cite{ALICE502a} and protons \cite{ALICE_protons} from the nucleus using the Zero Degree Calorimeter. However, the emission of other charged particles remains unexplored so far.

Our research aims to prepare a comprehensive description of this mechanism, attempts to explain the existing ALICE data, and makes predictions for future experiments.

\section{Photon-induced particle emission}

\begin{figure}[!b]
\centering
         a) single, 1$\gamma$-ray\quad\quad\quad b) single, 2$\gamma$-rays \quad\quad\quad c) mutual, 1$\gamma$1$\gamma$-rays\\
       \includegraphics[width=0.32\textwidth]{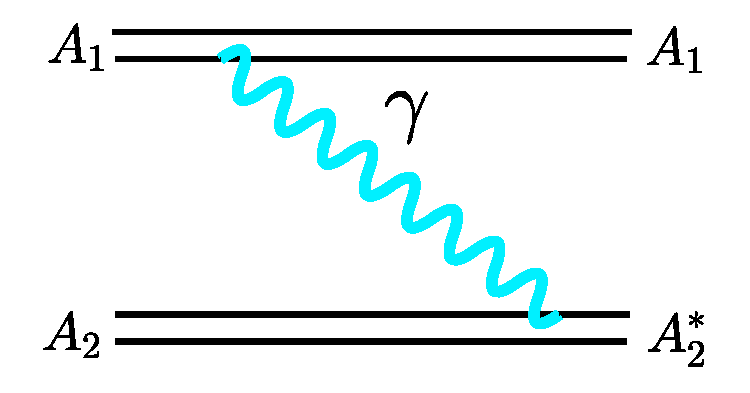}
        \includegraphics[width=0.32\textwidth]{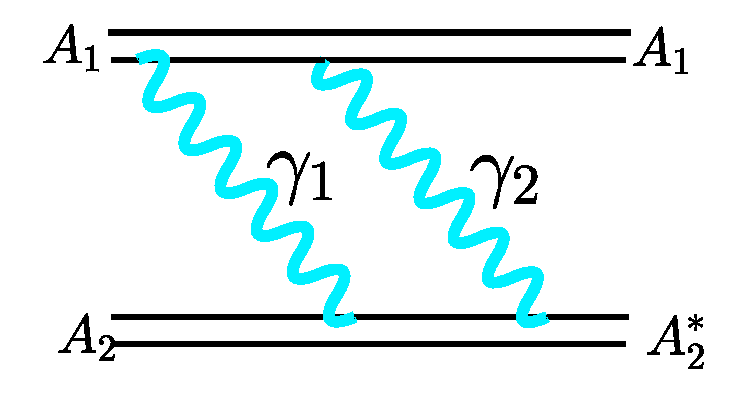}
        \includegraphics[width=0.32\textwidth]{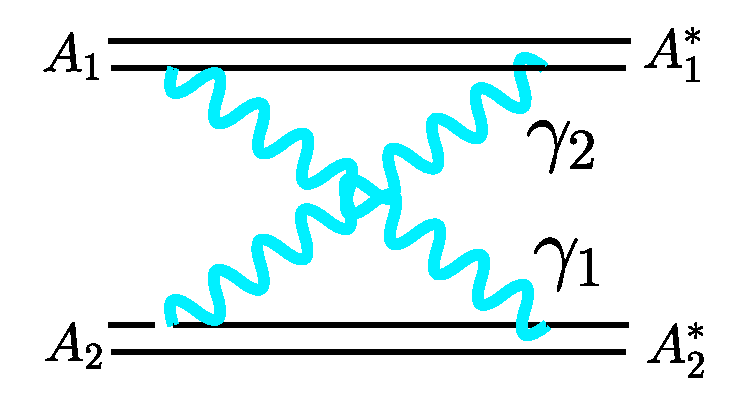}
    \caption{Nuclei excitation schemes in the UPC: (a), (b) single nucleus excitation (by various number of photons), (c) mutual nuclei excitations (multiple photon exchange).}
    \label{fig:feyn}\vspace{-0.2cm}
\end{figure}

%\section{Equivalent Photon Approximation}
The Coulomb interactions are depicted by one, two, three, or more photon exchanges, which can excite colliding nuclei. The examples of the scenario are presented in Fig.~\ref{fig:feyn}. Panels (a) and (b) - "single" show the situation when only one of nuclei, taking part in the collision, is excited by a given number of $\gamma$-rays exchanges; panel (c) displays mutual excitation of both participants. In our calculations up to 4 $\gamma$-rays exchanges are included.
\vspace{-0.cm}
%---------------------------------------------------------
%---------------------------------------------------------
 %   \vspace{-0.5cm}

The total cross section for $\gamma$-ray absorption can be written as: 
% Thus nuclei can be excited by single or mutual photon exchange.
% \vspace{-0.2cm}
\begin{eqnarray}
%\begin{split}
     \sigma_{tot} &=& \sigma_{single} + \sigma_{mutual},\\
    \sigma_{single} &=& \sigma^{(1\gamma)} + \sigma^{(2\gamma)} + \sigma^{(3\gamma)} + \dots, \\
    \sigma_{mutual} &=& \sigma^{(1\gamma1\gamma)} + \sigma^{(1\gamma2\gamma)} + \sigma^{(2\gamma1\gamma)} + \sigma^{(2\gamma2\gamma)}  + \dots .
%    \end{split}
\end{eqnarray}

The formula for the cross section contains photon flux ($N(E_\gamma,b)$), photoabsorption cross section ($\sigma_{abs}(E_\gamma)$) \cite{totabs} and the neutron emission probability density ($P_k(E_\gamma)$)
%\vspace{-0.2cm}

    \begin{equation}
    \sigma_{kX} = \int \int P_k(E_\gamma) \sigma_{abs}(E_\gamma) N(E_\gamma,b) 2\pi b db dE_\gamma.
    \label{eqsigma}
    \end{equation}
Above $k$ is a multiplicity of neutrons or protons ($X$). The $b$ is the impact parameter for the process.

The flux of photons $N(E_\gamma,b)$ is estimated by the Equivalent Photon Approximation (EPA) \cite{Klusek-Gawenda:2013ema} and the details are given in Ref.~\cite{jucha2024}.
    
     The excited nucleus cools down by emission of particles, photons and/or splitting into fragments in fission or fragmentation processes.
%\vspace{-0.3cm}
%\section{Photon-induced particle emission}
Among many existing afterburners, we choose the following models of nuclear deexcitation:
\begin{itemize}
    \item {\bf GEMINI++} - statistical model of nucleus deexcitation \cite{gemini} assumes full photon energy is transformed into the excitation energy of nucleus ($E_{exc}=E_{\gamma}$); 
%    \item {\bf TCM} - T \cite{jucha2024};
    \item {\bf HIPSE} - heavy-ion particle emission generator, pre-equilibrium physics is included \cite{hipse1} ($E_{exc}\neq E_{\gamma}$);
    \item {\bf EMPIRE} - a modular system of nuclear reaction codes for advanced nuclear reaction modeling using various theoretical models \cite{empire} ($E_{exc}\neq E_{\gamma}$).
\end{itemize}
In Ref.~\cite{jucha2024} we also proposed an alternative phenomenological approach - the two-component model (TCM) where the probability density of GEMINI++ is corrected as 
\begin{equation}
P(E_{exc};E_{\gamma})=c_1(E_{\gamma}) \delta \left( E_{exc}-E_{\gamma} \right)+c_2(E_{\gamma}) / E_{\gamma}.
\end{equation}
The $c_1$ and $c_2$ are functions of $E_{\gamma}$, parametrized with one free parameter.
This model gives a reasonable neutron multiplicity cross section but needs further development.
\begin{figure}[!b]
    \centering
        \includegraphics[width=0.8\textwidth]{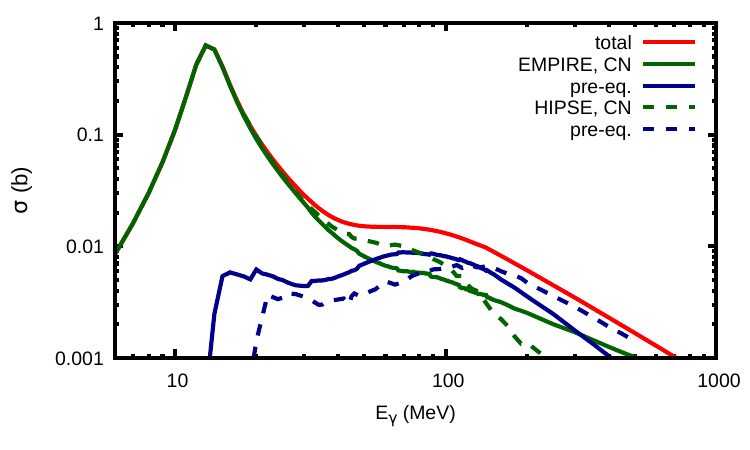}
    \caption{The cross section of the pre-equilibrium emission (pre-eq.) compared to compound nucleus production (CN) extracted from the HIPSE (n+$^{207}$Pb) and EMPIRE ($\gamma$+$^{208}$Pb) calculations.}
    \label{fig:pre-eq}
\end{figure}
%---------------------------------------------------------

\begin{figure}[!bt]
    \centering
 \setlength{\unitlength}{0.2\textwidth}
\begin{picture}(5,4.5)
\put(0.8,2.2){\includegraphics[width=0.8\textwidth]{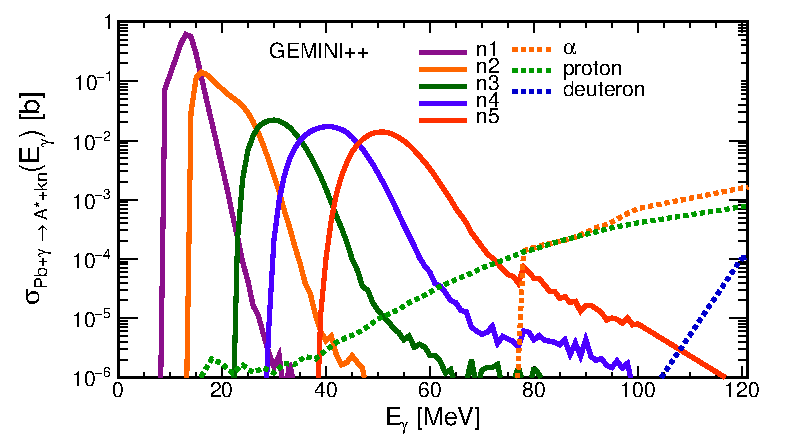}}
\put(0.7,4.1){a)} 
 \put(0.8,0.){\includegraphics[width=0.8\textwidth]{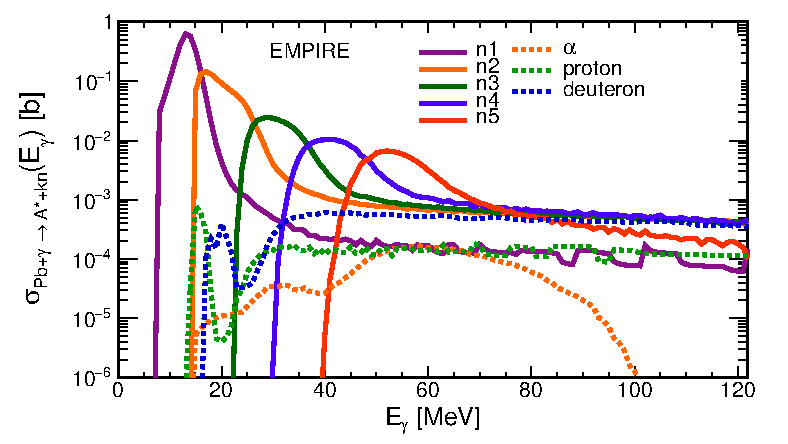}}
 \put(0.7,2.1){b)}  
\end{picture}
%        c) \includegraphics[width=0.8\textwidth]{hipse_prob.eps}
    \caption{The cross section of emission: 1-5 neutrons, 1 proton, 1 deuteron and 1 $\alpha$-particle (separately) estimated by GEMINI++ (a) and EMPIRE (b) approaches.}
    \label{fig:cross_sec}
\end{figure}

Fig.~\ref{fig:pre-eq} shows the onset of pre-equilibrium emission estimated in the HIPSE and EMPIRE approaches. For photon energies above 30~MeV the photon energy is dissipated by the emission of particles before thermal equilibration of the nucleus. Therefore, only a part of $E_{\gamma}$ is transformed into $E_{exc}$ of the nucleus. 
Furthermore, the prompt emission of particles (neutrons, protons and deuterons) changes the mass and charge of the thermal equilibrated nucleus. Following EMPIRE estimation, for $E_{\gamma}>$~50~MeV (for HIPSE this limit is two times higher), the initial $^{208}$ Pb loses at least one neutron. It has to be taken into consideration applying the de-excitation procedure.
Moreover, Fig.~\ref{fig:cross_sec} shows the minimum photon energy $E_{\gamma}$ necessary to emit a given multiplicity of neutrons or charged particles. In statistical code GEMINI++ it is very improbable to get a channel of 1p1n, as the phase space for this reaction is very small. The EMPIRE and other codes, where pre-equilibrium emission provides long $E_{\gamma}$ tails (see Fig.~\ref{fig:cross_sec} (b)) for the 1n channel, assure also that 1p1n and 1p2n de-excitation are possible.

\begin{figure}[!bt]
    \centering
        \includegraphics[width=0.8\textwidth]{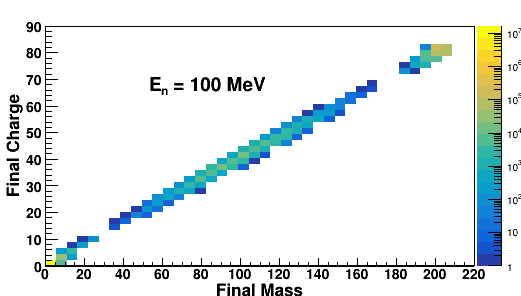}
    \caption{ Final mass-charge distribution of $n+^{207}Pb\rightarrow A^*+X$ at $E_{n}$=100~MeV reaction obtained with HIPSE+GEMINI++ model.}
    \label{fig:prefrag}
\end{figure}

The final mass-charge distribution calculated in the HIPSE + GEMI ++ model, presented in Fig.~\ref{fig:prefrag} gives the idea that apart from emission of particles and $\gamma$-rays, the fission process of the photon-induced Pb nuclei is also possible. Unfortunately, HIPSE cannot calculate the photon-induced reactions, but neutron-induced one, is quite a good approximation of the $\gamma+Pb\rightarrow A^*+X$ process.  Apart from evaporation residues, accompanied
by light charge nuclei (with mass range A$>$180 and A$<$40) there is a central part illustrating the noticable contribution of the fission phenomena. This is confirmed by the fission cross section presented in Table~\ref{tab5}.

%---------------------------------------------------------
%---------------------------------------------------------------
\begin{table}[!bt]
\caption{Total cross sections (in barn) for neutrons and charged particle emission in UPC $^{208}$Pb+$^{208}$Pb with collision energy $\sqrt{s_{NN}}$ = 5.02~TeV calculated with pure GEMINI++, HIPSE+GEMINI++ and EMPIRE. The results for inclusive channels kn and 1pXn, 1dXn and 1$\alpha$Xn are compared with the ALICE data, Ref.~\cite{ALICE502a,ALICE_protons}.
}
\centering
\begin{tabular}{c|c|c|c|c}
\hline
$\sigma$ [b] &GEMINI++ & HIPSE+GEMINI++ & EMPIRE & ALICE data \\
  \hline 
  1n & 90.4 &124.1 & 100.8 &  108.4 $ \pm\, 3.9$\,\,\,\,\, \\
  2n & 24.9 & 15.6 &  25.3 &  25.0 $\pm\, 1.3$\,\,\, \\
  3n & 3.3 &  4.9 &   5.9 &  7.95 $\pm\, 0.25$ \\
  4n & 2.4 &  3.6 &   5.6 &  5.65 $\pm\, 0.33$ \\
  5n & 1.5 &  3.9 &   3.4 &  4.54 $\pm\, 0.44$ \\
  \hline
  n$_{tot}$ & 122.6  &151.6 & 152.2 &151.5 $\pm\, 4.7$\,\,\,\,\, \\
  \hline
  1p &19.6 &28.5 &7.1 & 40.4 $\pm\, 1.7$\,\,\,\\
  1$\alpha$ & 30.7 & 42.4 &64.0 & - \\
  1d &6.91 & 5.2 & 2.7 & - \\
   fission &  10.8 &  18.3 & - & - \\ \hline
\end{tabular}
\label{tab5}
\end{table}

The cross section of the production of a given number of neutrons or charged particles in the ultraperipheral reaction $^{208}$Pb+$^{208}$Pb with collision energy $\sqrt{s_{NN}}$ = 5.02~TeV calculated with pure GEMINI++, HIPSE + GEMINI++ and EMPIRE is shown in Table~\ref{tab5}. The ALICE experimental data \cite{ALICE502a,ALICE_protons} are approximately reproduced by HIPSE+GEMINI++ and EMPIRE models. The uncertainty of theoretical estimations is less than 10\% and comes mainly from statistics (up to $10^6$ events for each $E_{\gamma}$) and the integration procedure in Eq.~\ref{eqsigma}. More details are presented in Ref.~\cite{jucha2024}.

\section{Conclusion}
%\begin{itemize}
   % \item 
    For a photon energy larger than 50 MeV only part of the energy is transferred directly to the equilibrated, excited nucleus.
   % \item 
%    Increasing the number of photons in single and mutual exchange decreases the impact of neutron emission cross section correction.
Multiphoton exchanges increase the cross section for neutron emission.
    %\item 
    Mutual excitations give less than 1\% of a single excitation of neutron emission cross section in the UPC.
   % \item 
    In addition to neutrons, charged particles are also emitted and protons can be measured in the proton Zero Degree Calorimeter. Our calculation gives reasonable reproduction of the ALICE data (see \cite{ALICE502a,ALICE_protons}).
Here we have focused on $\gamma+Pb\rightarrow A^*+X$ process which is an ingredient for the UPC collisions. We have shown the role of preequilibrium within EMPIRE and HIPSE approaches and corresponding light particle multiplicity distribution.

%\end{itemize}

%uncomment the following lines to place a figure
%\begin{figure}[htb]
%\centerline{%
%\includegraphics[width=12.5cm]{Fig1}}
%\caption{Plot of ...}
%\label{Fig:F2H}
%\end{figure}
\bibliography{biblio}
\bibliographystyle{unsrt}

\end{document}